\documentclass[aps,prd,twocolumn,showpacs,superscriptaddress,nofootinbib,preprintnumbers,floatfix]{revtex4}

\usepackage{amsmath}
\usepackage{amssymb}
\usepackage{hyperref}
\hypersetup{colorlinks=True,citecolor=blue}
\usepackage{graphicx}
\usepackage{bbm}
\usepackage{dcolumn}
\usepackage{enumitem}
\usepackage{color}

\begin{document}

\title{A novel approach to reconstructing signals  of isotropy violation  \\from a masked CMB sky}

\author{Pavan K. Aluri}
\affiliation{IUCAA, Pune - 411007, India}
\author{Nidhi Pant}
\affiliation{IUCAA, Pune - 411007, India}
\author{Aditya Rotti}
\affiliation{IUCAA, Pune - 411007, India}
\affiliation{Florida State University, Tallahassee, FL 32304, USA}
\author{Tarun Souradeep}
\affiliation{IUCAA, Pune - 411007, India}

\begin{abstract}
Statistical isotropy (SI) is one of the fundamental assumptions made in cosmological model building. This assumption is now being rigorously tested using the almost full sky measurements of the CMB anisotropies. A major hurdle in any such analysis is to handle the large biases induced due to the process of masking. We have developed a new method of analysis, using the bipolar spherical harmonic basis functions, in which we semi-analytically evaluate the modifications to SI violation induced by the mask. The method developed here is generic and can be potentially used to search for any arbitrary form of SI violation. We specifically demonstrate the working of this method by recovering the Doppler boost signal from a set of simulated, masked CMB skies.
\end{abstract}


\maketitle

\section{Introduction}

Cosmic microwave background (CMB) observations have ushered in the precision era in cosmology. This enables us to test the basic premises on which the standard cosmology rests, such as the assumption of statistical isotropy via the \emph{Cosmological Principle}. Some claims of statistical isotropy (SI) breakdown, that appeared in CMB literature were revisited by WMAP and PLANCK satellite science teams; see Ref.~\cite{WMAP7anom,Planck23} and the references there in.

The Bipolar Spherical Harmonic (BipoSH) formalism \cite{Hajian03,Joshi12a} is a powerful tool to identify and study statistical isotropy (SI) violation. This has been one of the methods used to search for deviation from isotropy in the CMB data \cite{Planck23, WMAP7anom, AH-TS-NC}.

The idea behind searching for SI violation is simple : it involves searching for non-vanishing BipoSH coefficients. This is made non-trivial due to the presence of a mask, as it induces spurious correlations that aliases the measurements. In this article, we  present a new methodology based on the BipoSH formalism, to estimate signals of SI violation from a masked CMB sky.  The algorithm presented here is generic, and can be potentially applied to recover any signal of SI violation, such as weak lensing, hemispherical power asymmetry etc. The Doppler boost is a known isotropy violation in the  CMB at a challenging level of subtlety. In this article we demonstrate the recovery of the Doppler boost field from a Doppler boosted CMB sky in the presence of a mask using our formalism. 

The paper is structured as follows. We briefly recap the BipoSH formalism in Sec.~\ref{sec:biposh}. In Sec.~\ref{sec:anisoeffect} we first discuss the SI violation induced due to Doppler boosting and masking separately and finally the resultant SI violation when both the effects are considered together. In section \ref{sec:partskyestmtr} we discuss the minimum variance estimator that takes into account the effects of masking for an arbitrary signal. The performance of the new estimator is discussed in Sec.~\ref{sec:demo}, with the specific example of Doppler boost. Finally our conclusions are presented in Sec.~\ref{sec:concls}.

\section{BipoSH : A brief summary}
\label{sec:biposh}

A CMB anisotropy map, $T (\hat{n})$, defined on a sphere is conventionally decomposed in terms of spherical harmonics $(Y_{lm})$ as,
\begin{equation}
T (\hat{n}) = \sum_{l=1}^{\infty} \sum_{m=-l}^{+l} a_{lm} Y_{lm} (\hat{n})\,,
\label{eq:spharmexp}
\end{equation}
where $\hat{n}$  denotes the position on the sphere, $T (\hat{n})$ are the temperature anisotropies observed in the direction $\hat{n}$ and $a_{lm}$ are the spherical harmonic coefficients of expansion.

The two-point correlation function is defined as,
\begin{equation}
C(\hat{n}_1,\hat{n}_2) = \langle T(\hat{n}_1) T(\hat{n}_2) \rangle\,,
\label{eq:2ptfundef}
\end{equation}
where $\langle \cdots \rangle$ denotes an in principle average over an ensemble of statistically independent CMB realizations. If the field $T$ is statistically isotropic, then it can be argued that the correlation function
cannot have any explicit directional dependence, and that it can only be a function of the separation $\theta$ between the two directions $\hat{n}_1$ \& $\hat{n}_2$, where $\cos \theta = \hat{n}_1 \cdot \hat{n}_2$. As a result of this simplification, the two-point correlation function is given by,
\begin{equation}
C(\hat{n}_1,\hat{n}_2) \equiv C(\theta) = \frac{1}{4\pi} \sum_{l=1}^{\infty} (2l+1) C_l P_l(\cos\theta)\,,
\end{equation}
where $P_l$ are the Legendre polynomials and $C_l$ is the well known \emph{angular power spectrum}. When expressed in terms of the harmonic space covariance matrix,
\begin{equation}
\langle a_{lm} a_{l'm'} \rangle = C_l \delta_{ll'} \delta_{mm'}\,, \label{isocovar}
\end{equation}
$C_l$'s are the diagonal and the only non-vanishing elements of the matrix. Hence for a statistically isotropic Gaussian random field, $C_l$ completely characterize the statistical properties of the field.

In the absence of statistical isotropy, $C_l$ do not completely describe the CMB sky and the Bipolar spherical harmonic (BipoSH) basis functions \cite{Hajian03} are better suited for this case. BipoSH form a complete orthonormal basis for an $S^2 \times S^2$ space, and hence can be used to describe any two point correlation function for a field defined on a sphere as
\begin{equation}
C(\hat{n}_1, \hat{n}_2) = \sum_{L, M, l_1, l_2} A^{LM}_{l_1l_2} \{ Y_{l_1}(\hat{n}_1) \otimes Y_{l_2}(\hat{n}_2) \}_{LM} \,,
\end{equation}
without making assumptions about statistical isotropy of the field. Here $A^{LM}_{l_1 l_2}$ are the coefficients of expansion and $\{ Y_{l_1} \otimes Y_{l_2} \}_{LM} = \sum_{m_1 m_2} \mathcal{C}^{LM}_{l_1 m_1 l_2 m_2} Y_{l_1 m_1} Y_{l_2 m_2}$ are the bipolar spherical harmonic basis functions \cite{varshalovich}. $C^{LM}_{l_1 m_1 l_2 m_2}$ are the Clebsch-Gordon coefficients whose indices satisfy the properties :
a)~$|l_1-l_2| \leq L \leq |l_1 + l_2|$,
b)~$m_1 + m_2 =M$,
c)~$-l_1 \leq m_1 \leq l_1$, and
d)~$-l_2 \leq m_2 \leq l_2$.

The harmonic space covariance matrix for a statistically anisotropic field, is fully described in terms of the BipoSH coefficients as,
\begin{equation}
\langle a_{l_1 m_2} a_{l_2 m_2} \rangle = \sum_{LM} A^{LM}_{l_1l_2} \mathcal{C}^{LM}_{l_1 m_1 l_2 m_2} \,, \label{anisocovar}
\end{equation}
in complete analogy with Eq.~[\ref{isocovar}]. Therefore the BipoSH coefficients completely characterize the statistical properties of a Gaussian, statistically anisotropic random field. It can be shown that $A^{00}_{ll}=(-1)^l \sqrt{2l+1} C_l$. Hence the BipoSH coefficients are a generalization of the commonly studied angular power spectrum. Note that Eq.~[\ref{anisocovar}] can also be used to define an estimator for the BipoSH coefficients as follows,
\begin{equation}
\hat{A}^{LM}_{l_1 l_2} = \sum_{m_1 m_2} \mathcal{C}^{LM}_{l_1 m_1 l_2 m_2} a_{l_11m_1 }a_{l_2 m_2}\,.
\end{equation}
It is easy to see that this is an unbiased estimator, since it converges to the true BipoSH coefficients on averaging over an ensemble of CMB skies. These BipoSH coefficients have the following symmetry properties,
\begin{itemize}
 \item Exchange symmetry : $C(\hat{n}_1, \hat{n}_2)= C(\hat{n}_2, \hat{n}_1)$
 \begin{equation}
   A^{LM}_{l_2 l_1} = (-1)^{l_1+l_2-L} A^{LM}_{l_1 l_2} \,,
 \end{equation}
 \item Reality condition : $C^*(\hat{n}_1, \hat{n}_2)= C(\hat{n}_1, \hat{n}_2)$
 \begin{equation}
 {A^{LM}_{l_1 l_2}}^* = (-1)^M (-1)^{l_1+l_2-L} A^{L,-M}_{l_1 l_2} \,.
 \end{equation}
\end{itemize}
Alternatively, following the triangularity condition between \{$l_1,l_2,L$\},
it is convenient to define the BipoSH coefficients as $A^{LM}_{l,l+D}$, where we set $l_1=l$, and $l_2=l+D$, with the constraint that $|D| \leq L$. Needless to say, following the above mentioned properties of the BipoSH coefficients, determining the BipoSH coefficients for  $0 \leq D \leq L$ and $0 \leq M \leq L$ will amount to computing all the BipoSH coefficients, $A^{LM}_{l_1l_2}$.

BipoSH have been a powerful tool in exploring isotropy violation in a blind as well as honed manner \cite{Kumar14}. It has earlier been used to probe SI violation as induced due to specific models of isotropy violation (phenomenological), cosmic topology (cosmological) and beams (systematic) \cite{WMAP7anom,Planck23,Ghosh07,Joshi10,Joshi12b}.


\section{Effect of isotropy violating phenomena}
\label{sec:anisoeffect}

Here we analytically derive the form of BipoSH coefficients as a function of multipole $l$, induced by Doppler boosting of the CMB sky. Following this, we derive the effect of masking on BipoSH coefficients, in addition to the anisotropy due to Doppler boost. Based on these analytic studies, a new estimator is obtained to recover the Doppler boost vector from a masked CMB sky.


\subsection{BipoSH coefficients due to Doppler boosting of CMB anisotropies}

One of the known phenomena that leads to breakdown of statistical isotropy is our relative motion ($\vec{v}$) with respect to the CMB rest frame (i.e. the frame in which an observer does not see a dipole). The strongest anisotropy in an otherwise uniform background temperature of $\mathcal{T}_0=2.7255$~Kelvin \cite{Mather1999,Fixsen09}, is an excess (deficit) temperature in the direction (opposite direction) of this relative motion. Doppler boost ($\vec{\beta}=\vec{v}/c$) that induces this large scale dipole temperature anisotropy (=$\mathcal{T}_0\vec{\beta}\cdot\hat{n}$) has a well measured amplitude and direction given by
$|\vec{\beta}|=1.23 \times 10^{-3}$ and $\hat{\beta}=(264^\circ,48^\circ)$ in galactic co-ordinates, respectively \cite{Kogut1993,Fixsen1996}. In addition to this dipole anisotropy, the velocity boost also modulates and aberrates the CMB temperature fluctuations leading to SI violation \cite{Challinor02, Amendola11, Chluba11, Kosowsky11, Notaria12, Suvodip14a}. These aberration and modulation effects can be used to make an independent measurement of the velocity of our local motion. It has been measured with ESA's PLANCK satellite data also \cite{Planck27,Adhikari15}. 

Doppler boosted CMB temperature anisotropies are given by \cite{Planck27},
\begin{equation}
T(\hat{n}) = \mathcal{T}_0 \vec{\beta} \cdot \hat{n} + \Theta \left(\hat{n}-\nabla(\vec{\beta}\cdot \hat{n})\right) \left(1+b_\nu\vec{\beta} \cdot \hat{n} \right) \,,
\label{eq:dbansio}
\end{equation}
where $T$ and $\Theta$ correspond to boosted and unboosted temperature anisotropies respectively, and $\vec{\beta} =\vec{v}/c$ is the Doppler boost vector. From Eq.~[\ref{eq:dbansio}] we see that, in addition to the dominant dipole component due to the Doppler effect, Doppler boosting also aberrates and modulates the CMB anisotropies. The modulation component has a frequency dependent factor $b_\nu=(\nu/\nu_0)\coth(\nu/2\nu_0)-1$, where $\nu_0=k_B \mathcal{T}_0/h \approx 57$~GHz.

For the rest of the discussion we will not bother with the first term in Eq.~[\ref{eq:dbansio}] since we often work with dipole subtracted maps. Evaluating the two point correlation function for the temperature anisotropies of a Doppler boosted CMB sky described in Eq.~[\ref{eq:dbansio}] and ignoring the dipole term, it can be shown that the resultant BipoSH coefficients can be cast in the following form,
\begin{equation}
A^{LM}_{l_1 l_2} = \left(A^{LM}_{l_1 l_2}\right)_{ub.cmb} + \beta_{LM} H^{L}_{l_1 l_2} \,,
\label{eq:biposh-doppler}
\end{equation}
up to first order in the isotropy violating Doppler boost field, $\beta (\hat{n}) = \vec{\beta} \cdot \hat{n}$. Note that Doppler boosting generates only the $L=1$ BipoSH modes.
The term $\left(A^{LM}_{l_1 l_2}\right)_{ub.cmb}$ corresponds to BipoSH coefficients due to \emph{unboosted CMB} anisotropies, $\beta_{LM}$ are the spherical harmonic coefficients of  $\beta(\hat{n})$, and $H^{L}_{l_1 l_2}$ is the shape function corresponding to Doppler boosting given by,
\begin{equation}
H^{L}_{l_1 l_2} = b_\nu \left(G^{L}_{l_1 l_2}\right)_{mod} - \left(G^{L}_{l_1 l_2}\right)_{abr} \,,
\label{fssf}
\end{equation}
where,
\begin{subequations}
\begin{eqnarray}
\left(G^{L}_{l_1 l_2}\right)_{mod} &=& \frac{C_{l_1} + C_{l_2}}{\sqrt{4\pi}}
\frac{\Pi_{l_1}\Pi_{l_2}}{\Pi_{L}} C^{L 0}_{l_1 0 l_2 0} \,, \\
\left(G^{L}_{l_1 l_2}\right)_{abr} &=& \frac{\left[ C_{l_1} F(l_1,L,l_2) + C_{l_2} F(l_2,L,l_1) \right] }{\sqrt{4\pi}} \nonumber \\
 && \times \frac{\Pi_{l_1}\Pi_{l_2}}{\Pi_{L}} C^{L 0}_{l_1 0 l_2 0} \,, \\
F(l_1,L,l_2) &=& \frac{l_1(l_1 + 1) + L(L + 1) - l_2(l_2 + 1)}{2} \,,
\end{eqnarray}
\end{subequations}              
where $\Pi_l = \sqrt{2l+1}$ and, $\left(G^{L}_{l_1 l_2}\right)_{mod}$ and $\left(G^{L}_{l_1 l_2}\right)_{abr}$ are the shape functions due to modulation and aberration effects, respectively and are completely specified by the angular power spectrum $C_l$.

A full sky minimum variance estimator for the velocity boost can then be defined as a weighted linear combination of the observed BipoSH coefficients, $A^{LM}_{l_1 l_2}$. See Appendix~\ref{sec:fsestmtr} for details.


\subsection{Effect of masking on BipoSH coefficients}

Even though an ideal CMB sky is isotropic, masking the sky to avoid foregrounds makes it highly anisotropic. In this section we derive the BipoSH coefficients arising due to the process of masking. Let a masked CMB sky be defined as,
\begin{equation}
\tilde{T}(\hat{n}) = \mathcal{W}(\hat{n}) T(\hat{n}) \,,
\label{eq:maskedaniso}
\end{equation}
where $\tilde{T}$ and $T$ denote the masked and unmasked CMB sky respectively and $\mathcal{W}(\hat{n})$ denotes the mask used for the analysis. Evaluating the two point correlation function for the masked CMB sky, it can be shown that the resultant BipoSH coefficients are given by, 
\begin{eqnarray}
\tilde{A}^{LM}_{l_1 l_2} &=& \sum_{l_3 l_4} \frac{\Pi_{l_3} \Pi_{l_4}}{\sqrt{4\pi}} \sum_{l_5 l_6} \frac{\Pi_{l_5} \Pi_{l_6}}{\sqrt{4\pi}} C^{l_1 0}_{l_3 0 l_5 0} C^{l_2 0}_{l_4 0 l_6 0} \nonumber \\
&& \times
\sum_{L' M' J K}
             \left\{
             \begin{array}{c c c}
              L  & l_1 & l_2 \\
              L' & l_3 & l_4 \\
              J  & l_5 & l_6
             \end{array}
             \right\} \nonumber \\
&& \times \,\, \Pi_{L'} \Pi_{J}
A^{L'M'}_{l_3 l_4} W^{J K}_{l_5 l_6}  C^{L M}_{L' M' J K} \,,
\label{eq:biposh:mask-unmask}
\end{eqnarray}
where $\tilde{A}^{LM}_{l_1 l_2}$ denotes the BipoSH coefficient for masked CMB sky, $A^{L'M'}_{l_3 l_4}$ denote the BipoSH coefficients of the full sky map, which is not assumed to be statistically isotropic for generality, $W^{J K}_{l_5 l_6}$ are the BipoSH coefficients of the mask and $\{\,\}_{3 \times 3}$ denotes the $9j-$symbol. The mask BipoSH coefficients are defined in the same way as the CMB sky. This equation is a generalization of the MASTER kernel \cite{Hivon2002} which describes how the angular power spectrum ($C_l$) is modified by mask. In the special case of $L=0$ and $L'=0$, this equation exactly reduces to the MASTER equation relating full sky power spectrum, $C_l$, to partial sky power spectrum, $\tilde{C}_l$.


\subsection{BipoSH of a masked anisotropic sky}

The full sky BipoSH coefficients of an anisotropic Doppler boosted CMB sky are given by Eq.~[\ref{eq:biposh-doppler}]. Thus the BipoSH coefficients of a masked Doppler boosted CMB sky can be obtained by substituting Eq.~[\ref{eq:biposh-doppler}] in Eq.~[\ref{eq:biposh:mask-unmask}] to get,
\begin{equation}
\tilde{A}^{LM}_{l_1l_2} = \left(\tilde{A}^{LM}_{l_1l_2}\right)_{ub.cmb} + \sum_{L'M'} \beta_{L'M'} K^{L'M'}_{LMl_1l_2} \,,
\label{eq:biposh:mask-unmask-aniso}
\end{equation}
where,
\begin{eqnarray}
\left( \tilde{A}^{LM}_{l_1 l_2} \right)_{ub.cmb} = (-1)^{l_1+l_2+L} \sum_{l_3} (-1)^{l_3} C_{l_3} \frac{\Pi_{l_3}^2}{\sqrt{4\pi}}
\nonumber \\
 \times \sum_{l_5 l_6}\frac{\Pi_{l_5} \Pi_{l_6}}{\sqrt{4\pi}}
 C^{l_1 0}_{l_3 0 l_5 0} C^{l_2 0}_{l_3 0 l_6 0} W^{L M}_{l_5 l_6}
     \left\{
      \begin{array}{c c c}
       l_5 & l_6 & L \\
       l_2 & l_1 & l_3
      \end{array}
     \right\} \,,
\end{eqnarray}
denotes the BipoSH coefficients generated due to masking an unboosted CMB sky characterized by $C_l$. $\beta_{L'M'}$ are the harmonic coefficients of the Doppler field $\beta(\hat{n})=\vec{\beta}\cdot\hat{n}$, and  the term $\{\}_{2 \times 3}$ denotes the $6j-$symbol. The modified shape function (MSF), $K^{L'M'}_{LMl_1l_2}$, of the Doppler signal, is the masked analogue of the full sky shape function, $H^{L}_{l_1 l_2}$, in Eq.~[\ref{fssf}] and is given by,
\begin{eqnarray}
K^{L'M'}_{LMl_1l_2} = \sum_{l_3 l_4} \Pi_{L'} H^{L'}_{l_3 l_4} \frac{\Pi_{l_3} \Pi_{l_4}}{\sqrt{4\pi}} \sum_{l_5 l_6} \frac{\Pi_{l_5} \Pi_{l_6}}{\sqrt{4\pi}} C^{l_1 0}_{l_3 0 l_5 0}  \nonumber \\
\times C^{l_2 0}_{l_4 0 l_6 0}
\sum_{J K}
             \left\{
             \begin{array}{c c c}
              L  & l_1 & l_2 \\
              L' & l_3 & l_4 \\
              J  & l_5 & l_6
             \end{array}
             \right\}
        \Pi_{J} W^{J K}_{l_5 l_6}  C^{L M}_{L' M' J K} \,.
\label{eq:mixingkernel}
\end{eqnarray}
The modified shape function $K^{L'M'}_{LMl_1l_2}$ incorporates the mixing of modes $\{JK\}$ due to mask  and the intrinsic anisotropic modes \{$L'M'$\}, giving rise to the observed modes \{$LM$\}. This coupling is captured by the Clebsch-Gordan coefficient $C^{L M}_{L' M' J K}$ and the $9j-$symbol. 

Hence from Eq.~[\ref{eq:biposh:mask-unmask-aniso}], it can be seen that masking may leak power from an intrinsic anisotropic mode, $L'$, to any observed mode, $L$. Further note that the modified shape function due to masking is now phase ($M$,$M'$) dependent unlike in Eq.~[\ref{fssf}].  Here we emphasize that the formalism discussed in this section is completely generic, since we assumed no particular form of isotropy violation.


\section{Partial sky estimator}
\label{sec:partskyestmtr}


\subsection{Approximations}
In general, masking leads to highly entangled modes as seen from Eq.~[\ref{eq:biposh:mask-unmask-aniso}] and Eq.~[\ref{eq:mixingkernel}]. We now specifically consider the case of BipoSH coefficients generated due to a Doppler boosted CMB sky. Though in a full sky Doppler boosted CMB sky the signal is only in $L'=1$ mode of BipoSH coefficients, the masked Doppler boosted sky can have power leaked to $L \neq 1$ modes. The mode coupling of intrinsic anisotropic signal ($L'=1$) with the mask indices ($J$), giving rise to the observed modes ($L$) are tabulated in Table~\ref{tab:mixing-of-indices}.
\begin{table}[!hb]
\centering
\begin{tabular}{  c | c  }
  \hline
  \multicolumn{2}{c}{Intrinsic anisotropic signal, $L'=1$} \\
  \hline
  \\
  Observed BipoSH index, & Mask BipoSH index, \\
  $L$ & $J=|L-L'|$ to $L+L'$ \\
  \hline
  \\
  $L=1$ & $J={\bf 0}, 1,{\bf 2}$ \\
  $L=2$ & $J=1,{\bf 2},3$ \\
  $L=3$ & $J={\bf 2},3,{\bf 4}$ \\
  \hline
\end{tabular}
\caption{Illustrated here are the observed BipoSH indices ($L$) that result from
the mixing of an intrinsically anisotropic signal in $L'=1$ with the mask
BipoSH indices ($J$). For a mask which is largely azimuthally oriented, the dominant mask modes contributing to the mixing kernel are highlighted in bold letters.}
\label{tab:mixing-of-indices}
\end{table}

We now argue, that even though there is coupling between different modes, it is reasonable to assume that only modes with $L'=L=1$ and $M=M'$, which we refer to as the diagonal approximation,  are sufficient to evaluate Eq.~[\ref{eq:biposh:mask-unmask-aniso}]. 

Most masks used in CMB analysis can at first order be approximated as a band along the galactic equator where the astrophysical emission from our own galaxy is the highest and hence have a significant azimuthal symmetry. It can be argued that for such a mask, the mixing kernel is primarily diagonal in $\{M,M'\}$, motivated by the fact that for a band mask, which has perfect azimuthal symmetry, the $\{M,M'\}$ coupling can be shown to be identically zero.
\begin{figure}[!ht]
\centering
\includegraphics[width=0.45\textwidth]{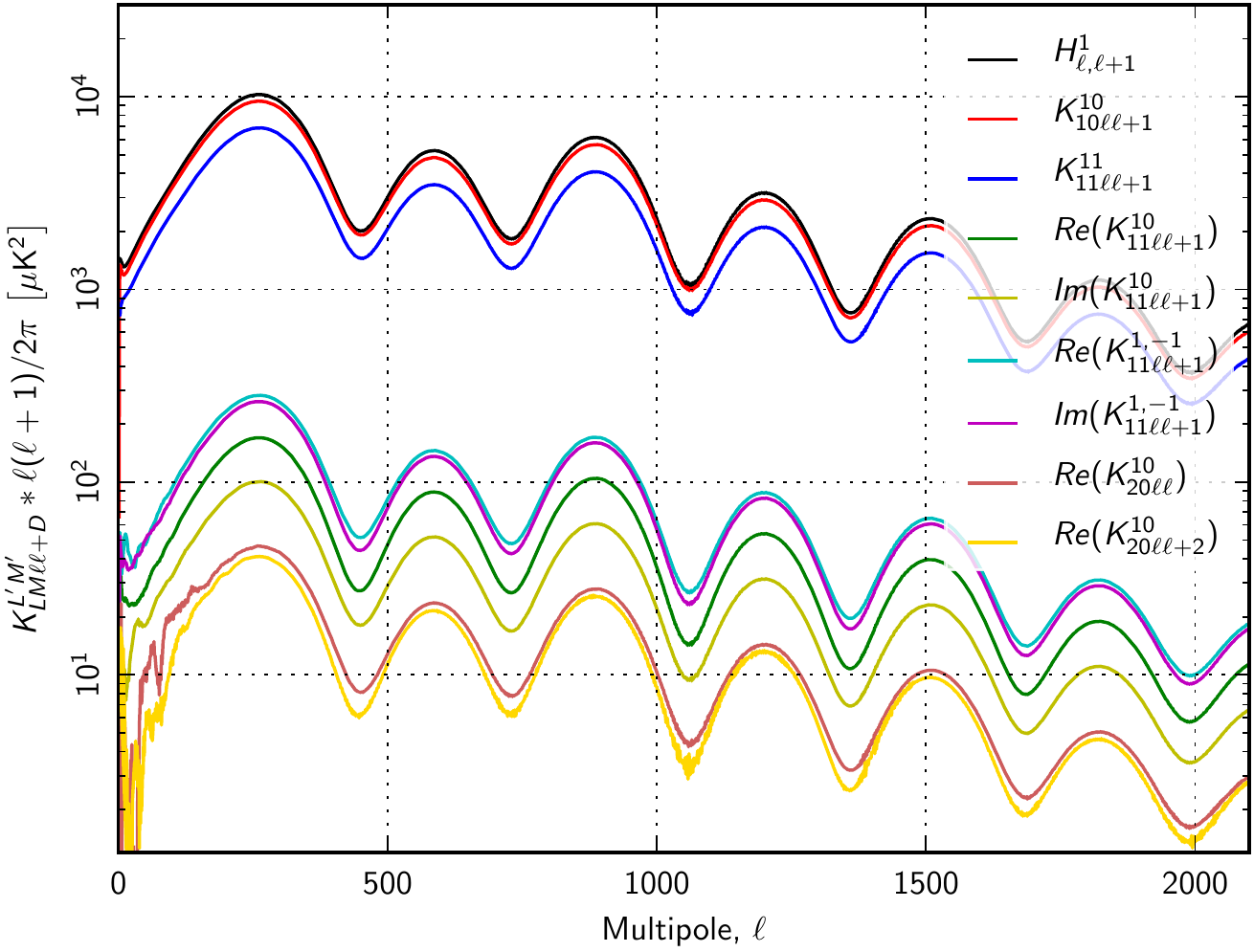}
\caption{In this figure we compare the full sky shape function $H^1_{l,l+1}$ and the modified shape function $K^{L',M'}_{L,M,l,l+1}$, modified due to mask, for the Doppler boosted CMB sky. The full sky shape function is independent of phase as it does not depend on $M$. The most dominant terms of the modified shape function is the diagonal denoted by  $K^{L,M}_{L,M,l,l+1}$, i.e. when $L=L'$ and $M=M'$. Also plotted are some of the off-diagonal terms of the shape function, i.e.  $L \neq L'$ and $M \neq M'$, which are seen to be negligible when compared to the diagonal and hence ignored in the analysis. Further note that for the diagonal terms, there is a $M$ dependent suppression of power as compared to $H^1_{l,l+1}$. The varying supression of power between modes corresponding to different $M$ is determined by the morphology of the mask.}
\label{fig:mask-fs-ps-shape}
\end{figure}
For mask BipoSH coefficients it is seen that the $J=0$ mode is the most dominant mode. Observing the various couplings in Table~\ref{tab:mixing-of-indices} it can be seen that only the $L=1$ mode couples to $L'=1$ mode via $J=0$, while all higher $L>1$ modes couple to $L'=1$ via $J>0$, suggesting that the contribution from terms where $L\neq L'$ will be sub-dominant.

We follow these heuristic arguments by quantitatively showing that the off-diagonal terms are infact subdominant by explicitly evaluating the kernel $K^{L'M'}_{LMll'}$. In Fig~[\ref{fig:mask-fs-ps-shape}] we compare the full sky shape function $H^L_{l\,l'}$ of Doppler boosting with the modified shape function, $K^{L'M'}_{LMll'}$, evaluated using the apodized version of the mask shown in Fig.~[\ref{fig:planck2015commonmask}] (used for all analyses presented in this article). Note that the diagonal terms are nearly two orders of magnitude larger than the off-diagonal terms. This justifies the approximations suggested above. 

Here it is also important to note that the modified shape function is different for different phase modes ($M=0,1$). This distinction is purely determined by the details of the mask. As we will see in the following sections, this difference in MSF corresponding to different phase modes ($M$), naturally compensates for the differential loss of signal in these modes while recovering the Doppler field harmonics.

Note that in Eq.~[\ref{eq:biposh:mask-unmask-aniso}], it is fairly complicated to algebraically solve for the Doppler field harmonics, $\beta_{LM}$, since it appears inside a summation. By making the approximations discussed in this section, we have simplified Eq.~[\ref{eq:biposh:mask-unmask-aniso}] to the form,
\begin{equation}
\tilde{A}^{L M}_{l_1 l_2} = \left(\tilde{A}^{LM}_{l_1 l_2}\right)_{ub.cmb} + \beta_{LM} K^{L M}_{L M l_1 l_2} \,.
\label{eq:biposh:mask-unmask-aniso-approx}
\end{equation}

This is particularly useful, since it can be used to define a Doppler estimator, similar to the full sky estimator.


\subsection{Weighted variance estimator using modified shape function (WV-MSFE)}
\label{app:WVMSFE}

In this section we briefly discuss the Doppler estimator.  Eq.~[\ref{eq:biposh:mask-unmask-aniso-approx}] can be inverted to arrive at an estimator for the Doppler field harmonics,
\begin{equation}
\hat{\beta}_{LM} = \frac{\tilde{A}^{L M}_{l_1 l_2} - \langle \tilde{A}^{LM}_{l_1 l_2}\rangle_{ub.cmb}}{K^{L M}_{L M l_1 l_2}} \,,
\end{equation}
where $\tilde{A}^{L M}_{l_1 l_2} $ denotes the BipoSH coefficients derived from the data maps, and $\langle \tilde{A}^{LM}_{l_1 l_2}\rangle_{ub.cmb}$ is the expected bias due to mask and spatially varying noise in an observed map. The expected bias is estimated from an ensemble of simulations which are not Doppler boosted. This naive estimator can be optimized by minimizing its variance arising due to cosmic variance and instrument noise. The derivation of the estimator is discussed in Appendix~\ref{sec:psestmtr}. The estimator used to reconstruct the Doppler boost vector is given by,
\begin{equation}
\hat{\beta}_{LM} = \sum_{l_1 l_2} \hat{w}^{L}_{l_1 l_2} \frac{\hat{\mathcal{A}}^{LM}_{l_1 l_2}}{K^{LM}_{L\,M\,l_1\,l_2}} \,,
\label{opt_est}
\end{equation}
where $\hat{\mathcal{A}}^{LM}_{l_1l_2} = \tilde{A}^{LM}_{l_1l_2} - \langle \tilde{A}^{LM}_{l_1 l_2} \rangle_{ub.cmb}$ are the bias corrected  BipoSH coefficients. The weights $ \hat{w}^{L}_{l_1 l_2}$ which minimize the variance are given by the expression, 
\begin{equation}
\hat{w}^{L}_{l_1 l_2} = \frac{1}{\sum_{M} \left(\frac{\hat{\sigma}^{LM}_{l_1 l_2}}{K^{LM}_{L M l_1 l_2}}\right)^2}
\left[
\sum_{l'_1 l'_2} \frac{1}{\sum_{M} \left(\frac{\hat{\sigma}^{LM}_{l'_1 l'_2}}{K^{LM}_{L M' l'_1 l'_2}}\right)^2}
\right]^{-1} \,,
\end{equation}
where,
\begin{equation}
 \left(\hat{\sigma}^{LM}_{l_1 l_2}\right)^2 = \langle |\tilde{A}^{LM}_{l_1 l_2}|^2 \rangle_{ub.cmb} -
     |\langle \tilde{A}^{LM}_{l_1 l_2} \rangle_{ub.cmb}|^2 \,,
\end{equation}
is the variance of unboosted map's BipoSH coefficients.

We see that the effective weights are $M$ dependent owing to the $M$ dependent shape function in Eq.~[\ref{opt_est}], unlike in the full sky estimator. Since the estimator is weighed by the modified shape function which accounts for loss of power due to mask, it naturally corrects for the reduced amplitude of the Doppler vector due to masking.


\section{Demonstration of the WV-MSF estimator}
\label{sec:demo}

We demonstrate the working of this newly proposed method for reconstructing the Doppler boost, by evaluating it on an ensemble of Doppler boosted simulations. We also run a parallel analysis on full skies to allow direct comparison of it's efficiency.


\subsection{Generating simulations}
\label{sec:sim}

We generated a set of 1000 simulations with characteristics of Planck 217~GHz instrument using the best fit theoretical $C_l$ corresponding to the cosmological parameters from Planck 2013 data \cite{Planck16}. The isotropic simulations were generated using the \texttt{synfast} facility of \texttt{HEALPix}\footnote{\url{http://healpix.sourceforge.net/}} \cite{Gorski05}. The Doppler boosted simulations are generated using the \emph{Code for Non-Isotropic Gaussian Sky} (\texttt{CoNIGS}) algorithm \cite{Suvodip14}. We injected Doppler boost with amplitude $|\vec{\beta}| = 1.23 \times 10^{-3}$, pointing towards the galactic coordinates $(l,b)=(264^\circ,48^\circ)$. Since these simulations are specific to the
217~GHz channel, we set the frequency dependent modulation factor to $b_\nu = 3$.

The noise simulations were generated separately and added to the CMB simulations.
In our analysis, we first worked with isotropic noise approximation and then with realistic spatially varying noise, so as to progressively increase the complexity of the data.
The isotropic noise simulations were generated using a white Gaussian noise spectrum corresponding to the 217~GHz instrument ($\theta_{fwhm}=5 ~\textrm{arcmin}$ \& $\sigma=4.8~\mu K/K$), as quoted in Planck Bluebook \cite{PlanckBlueBook}. To simulate the full mission (five surveys) noise levels, we divide the nominal (two surveys) noise standard deviation by $\sqrt{5/2}$.

The realistic, spatially varying noise maps were generated using the noise variance map (no cross-pixel correlations) corresponding to Planck 217~GHz channel available in the public archives\footnote{\url{http://irsa.ipac.caltech.edu/Missions/planck.html}}. 


\subsection{Mask and modified shape function}

\begin{figure}[tp]
 \centering
 \includegraphics[width=0.45\textwidth]{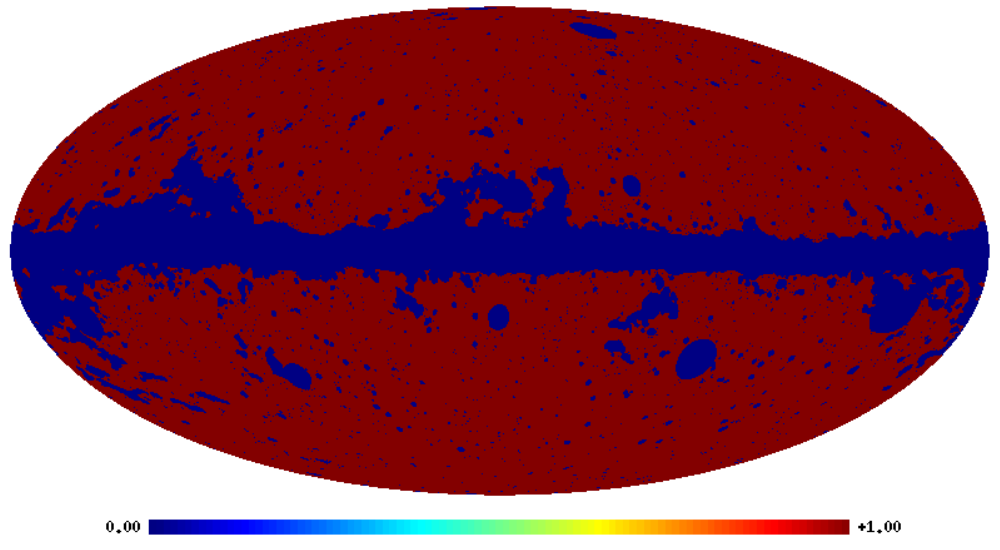}
 \caption{The unapodized galactic mask used in the analysis is shown here. This $\{1,0\}$ mask has an available sky fraction of $f_{sky} \approx 0.78$, which remains practically same on apodization. }
 \label{fig:planck2015commonmask}
\end{figure}

We use the common analysis mask used in the Planck analysis shown in Fig.~[\ref{fig:planck2015commonmask}]. Sharp \{0,1\} masks are not ideal for any band limited analysis as they result in heavy ringing at the edges of the mask. To avoid this unnecessary complication we apodize the mask with a Gaussian beam of 30 arcmin. The apodized mask has an effective sky fraction of $f_{sky} \approx 0.78$ which is practically same as the sky fraction available with the unapodized mask.

Given the apodized mask and the fiducial power spectrum used to generate the simulations, we numerically evaluate the modified shape function $K^{LM}_{LM l_1 l_2}$ using the expression in Eq.~[\ref{eq:mixingkernel}]. The $3j$ and $6j$ symbols are calculated using standard routines available in \texttt{SLATEC} numerical libray\footnote{\url{http://www.netlib.org/slatec/index.html}}. The modified shape functions evaluated and used in our analysis are depicted in Fig.~[\ref{fig:mask-fs-ps-shape}].
%


\subsection{Analysis and results}
\label{sec:analysis}

In this section we test the recovery of Doppler boost vector from simulated data using the new estimator presented in section~\ref{app:WVMSFE}. We perform the analysis using the full sky simulations in addition to the analysis on masked maps to allow direct comparison.

For the masked analysis, we first evaluate the ensemble averaged BipoSH coefficients and their variance from the unboosted, noise added, masked simulations. While the average is used to subtract the biases due to the mask and spatially varying noise, the variance is used in the evaluation of the minimum variance estimator in Eq.~[\ref{opt_est}]. The Doppler boost estimator is  evaluated on each realization of Doppler boosted simulations, yielding an ensemble of estimates of the Doppler field $\beta(\hat{n})=\vec\beta \cdot \hat{n}$. While the direction of the Doppler boost is determined using the \texttt{HEALPix} subroutine \texttt{remove\_dipole}, the Doppler boost amplitude  is recovered by first estimating the power in the reconstructed Doppler field :  $|\vec\beta| = 1.5 \sqrt{\beta_1/\pi}$ where $\beta_1 = \sum_M |\beta_{1M}|^2/3$. 

The Doppler power is not expected to vanish when estimated from an unboosted sky, and this bias is termed reconstruction noise. In order to have an unbiased estimate of the Doppler amplitude the reconstruction noise $\beta^N_1$ needs to be subtracted from the estimated Doppler power. This bias can be dealt with semi-analytically in the full sky case (see Appendix~\ref{sec:fsestmtr}). For the masked analysis we estimate the mean reconstruction noise by applying the Doppler estimator on a set of masked unboosted simulations. The unbiased estimate of the Doppler boost amplitude is given by $|\vec{\beta}| = 1.5 \sqrt{(\beta_1 - \beta^N_1)/\pi}$.
\begin{figure}[!ht]
  \centering
    \includegraphics[width=0.48\textwidth]{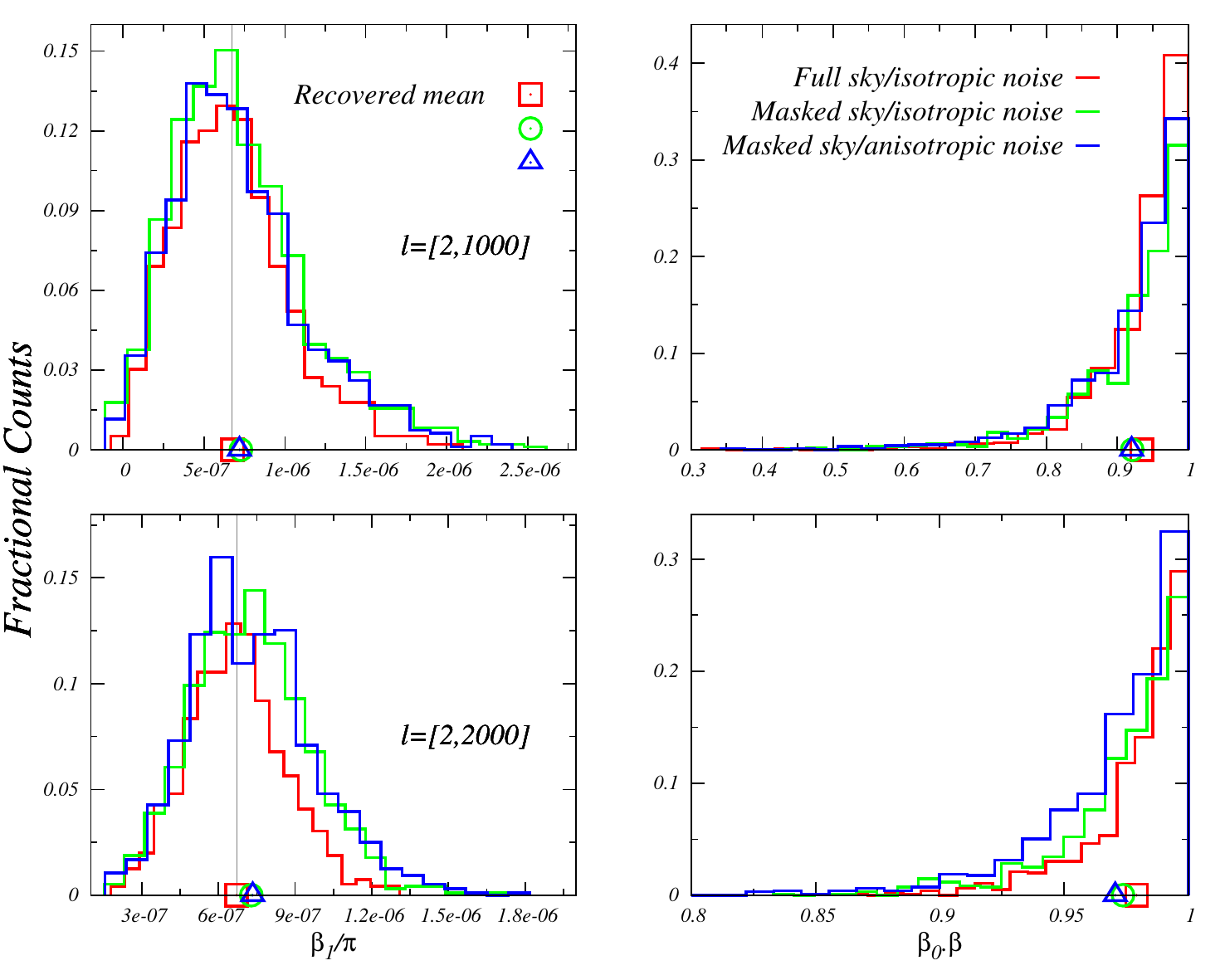}
\caption{This figure depicts the recovered Doppler boost parameters from analysis on simulated CMB skies masked with the apodized Planck common mask. While the left column depicts the histogram of the recovered Doppler power corrected for reconstruction noise bias, the right column depicts the histogram of the cosine of angular separation $\hat{\beta}_0 \cdot \hat{\beta}$, where  $\hat{\beta}_0$ and $\hat{\beta}$ denotes the injected and recovered  directions respectively. The top and bottom row denote the results from analysis in the multipole bins $l=[2,1000]$ and $[2,2000]$ respectively. The injected Doppler power is denoted by a vertical grey line.}
\label{fig:dopplerestm}
\end{figure}
The full sky analysis follows the same procedure as above, except that we don't have to bias subtract the BipoSH coefficients, since we only run the full sky analysis on simulations with isotropic noise. Note that one still needs to bias subtract the power in the Doppler field to arrive at the correct Doppler boost amplitude.
\begin{table}
\centering
\scriptsize
\begin{tabular}{ l c c c c c }
\hline 
Mask & $\Delta l$ & $|\beta|$ ($\times 10^{-3}$) & $b^\circ$ & $l^\circ$ \\
\hline
Full sky          & [2,1000] & 1.23 $\pm$ 0.33 & 46 $\pm$ 14    & 266 $\pm$ 26 \\
(isotropic noise) & [2,2000] & 1.23 $\pm$ 0.18 & 47.7 $\pm$ 8.2 & 266 $\pm$ 13 \\
                \hline
Common Mask       & [2,1000] & 1.28 $\pm$ 0.38 & 45 $\pm$ 15    & 262 $\pm$ 32 \\
(isotropic noise) & [2,2000] & 1.28 $\pm$ 0.20 & 46.6 $\pm$ 9.2 & 262 $\pm$ 14 \\
                \hline
Common mask         & [2,1000] & 1.27 $\pm$ 0.38 & 44 $\pm$ 15    & 263 $\pm$ 30 \\
(anisotropic noise) & [2,2000] & 1.28 $\pm$ 0.21 & 46.2 $\pm$ 9.2 & 262 $\pm$ 15 \\
\hline
\end{tabular}
\caption{The mean recovered Doppler amplitude and direction obtained from 1000 simulations are tabulated here. The boosted simulations were generated with a Doppler boost amplitude $|\vec{\beta}| = 1.23 \times 10^{-3}$ pointing towards the galactic coordinates $(l,b)=(264^\circ,48^\circ)$. The errors quoted are the standard deviation computed from probability density function inferred from the ensemble of reconstructed Doppler power and direction (l,b).}
\label{tab:doppler-power}
\end{table}

Finally we run the analysis on two multipole ranges, $[2,1000]$ and $[2,2000]$. At high multipoles $l \gtrsim 1000$, the results can be potentially affected by point source masks. By demonstrating robustness against a range of multipoles used in the analysis, we have shown that our results are not affected by the presence of point source masks.

Results of the reconstruction of Doppler boost vector are summarized in Fig.~[\ref{fig:dopplerestm}]. Table~\ref{tab:doppler-power} lists the mean recovered Doppler amplitude and direction from different multipole bins.  Note that the injected amplitude and direction are consistently recovered from the masked as well as full sky simulations with isotropic and anisotropic noise. We find that the error on the recovered Doppler amplitude and direction is larger when using smaller multipole range $l \in [2,1000]$ as compared to estimates from using the larger multipole range $l \in [2,2000]$ as expected. Similarly the error on estimates from masked sky is larger than the error on estimate from full sky, as expected due to the reduced sky fraction owing to the mask. Finally we reiterate that we don't have to perform any $f_{sky}$ corrections to the Doppler amplitude as is usually required by other methods.


\section{Conclusions}
\label{sec:concls}

For any reliable cosmological analysis, masking the CMB maps is inevitable to avoid biases due to galactic foregrounds. The effects of masking on the angular power spectrum are easy to reverse using the MASTER algorithm \cite{Hivon2002}.  

Here we developed a formalism similar to the MASTER algorithm but extended to all BipoSH coefficients which are a generalization of the well known angular power spectrum. Using this formalism, we derived an expression which describes how masking modifies an arbitrary isotropy violating signal. Though we write down the mask coupling matrix for the BipoSH coefficients, we don't try to invert the equation as done in the MASTER algorithm.  We simplified the equation using the symmetry properties of the mask and properties of Clebsch-Gordon coefficients.

As a very specific example, we studied how the Doppler boost signal is modified due to the presence of the mask. We showed that by making reasonable assumptions it is possible to cast the BipoSH coefficients of a masked Doppler boosted sky in a form similar to the BipoSH coefficients of Doppler boosted full CMB sky, where the full sky shape function is replaced by the modified shape function.
Eventually we obtained an estimator which recovers the Doppler signal from a masked CMB sky. This estimator naturally accounts for the loss of power due to masking, and hence does not require an additional $f_{sky}$ correction as required by other quadratic estimators.

We generated a set of simulations whose statistical properties are tailored to match the 217 GHz Planck maps. We used an apodized form of Planck 2015 common mask for our analysis. Finally we evaluated the newly derived estimator on masked CMB skies to demonstrate it's unbiased recovery of the injected signal. 

Finally we note that the method developed here are generic and can be used to reconstruct any arbitrary form of SI violation.

{\bf Acknowledgements} : We acknowledge the use of \texttt{HEALPix} package, and \texttt{SLATEC} library which is a part of the \texttt{NETLIB} software repository, that are freely available, in this work. We thank Suvodip Mukherjee for generating the Doppler boosted CMB realizations using \texttt{CoNIGS} for this work.



\appendix


\section{Full sky statistic}
\label{sec:fsestmtr}

Many isotropy violating signals can be written in the generic form \cite{Kumar14},
\begin{equation}
\hat{A}^{LM}_{l_1 l_2} = A^{LM}_{l_1 l_2} + \beta_{LM} H^L_{l_1 l_2} \,,
\label{generic_si_violation}
\end{equation}
where $\hat{A}^{LM}_{l_1 l_2} $ denote the BipoSH coefficients measured from the data, $A^{LM}_{l_1 l_2}$ denote the BipoSH coefficients of the statistically isotropic CMB sky,  $\beta_{LM}$ denote the parameters of the isotropy violating field, and $H^L_{l_1 l_2}$ denotes the spectral shape function induced by the particular form of isotropy violation.

For a statistically isotropic CMB field, $\langle A^{LM}_{l_1 l_2} \rangle = 0$
for $L>0$. Following Eq.~[\ref{generic_si_violation}], we can define a \emph{naive} estimator for the parameters $\beta_{LM}$ as follows,
\begin{equation}
\hat{\beta}_{LM} = \frac{1}{\sum_{l_1 l_2}}\sum_{l_1 l_2} \frac{\hat{A}^{LM}_{l_1 l_2}}{H^{L}_{l_1 l_1}} \,.
\end{equation}

However, the above estimator is not optimized to reduce the variance due to noise. It is possible to recover a signal with better signal to noise ratio by redefining the estimator as, 
\begin{equation}
\hat{\beta}_{LM} = \sum_{l_1 l_2} \hat{w}^L_{l_1 l_2} \frac{\hat{A}^{LM}_{l_1 l_2}}{H^{L}_{l_1 l_1}} \,,
\label{eq:estm_fs_mve}
\end{equation}
where the weights $\hat{w}^L_{l_1 l_2}$ are  to be chosen so as to minimize the quantity,
\begin{subequations}
\begin{eqnarray}
\hat{C}^{\beta \beta}_L &=& \langle \hat{\beta}_{LM} \hat{\beta}_{LM}^* \rangle \,, \\
&=& \sum_{l'_1 l'_2} (\hat{w}^{L}_{l'_1 l'_2})^2 \frac{2 C_{l'_1} C_{l'_2}}{(H^{L}_{l'_1 l'_2})^2} + C^{{\beta}{\beta}}_{L} \,, \label{eq:doppowerlc} \\
&=& N_{L} + C^{\beta\beta}_{L} \,, 
\end{eqnarray}
\end{subequations}
where $N_L$ is the reconstruction noise that is minimized by choosing appropriate weights, and $\hat{C}^{\beta \beta}_L$ \&  $C^{\beta\beta}_{L}$ denote the reconstructed and true power in the multipole $L$ respectively, of any isotropy violating field $\beta(\hat{n})$. Minimizing the reconstruction noise results in down weighting the noisy modes while giving more weightage to the signal dominated modes. In arriving at Eq.~[\ref{eq:doppowerlc}] we used the covariance of BipoSH coefficients given by \cite{Joshi12a},
\begin{equation}
\langle A^{LM}_{l_1 l_2} A^{L'M'^*}_{l'_1 l'_2}\rangle = 
C_{l_1} C_{l_2} \left[ \delta_{l_1 l'_1} \delta_{l_2 l'_2} + \delta_{l_1 l'_2} \delta_{l'_1 l_2}\right]\delta_{LL'} \delta_{MM'} \,,
\label{biposh-cov}
\end{equation}
which is valid when $L \neq 0$ and $L +l_1 +l_2$ is even.

This minimization of power is evaluated subject to the constraint $\sum_{l_1 l_2} \hat{w}^L_{l_1 l_2} = 1$, resulting in a constrained  minimization problem, which is solved using the method of Lagrange multipliers. The weights are determined by minimizing the function, 
\begin{equation}
\mathcal{L} = \hat{C}^{\beta \beta}_{L} -  \alpha \left[ \sum_{l_1 l_2} \hat{w}^{L}_{l'_1 l'_2} -1 \right] \,,
\label{eq:fslagrangemultiplier}
\end{equation}
where $\alpha$ is the Lagrange multiplier. Note that the weights for each BipoSH mode are assumed to be independent  of each other in the following sense,
\begin{equation}
\frac{\partial w^L_{l_1 l_2}}{\partial w^{L'}_{l_1' l_2'}} = \delta_{LL'}\delta_{l_1 l_1'}  \delta_{l_2 l_2'} \,. \label{ind-w}
\end{equation}
On setting the derivative of $\mathcal{L}$ with respect to the weights $\hat{w}^{L'}_{l'_1 l'_2}$ to zero and simplifying the resultant equation it can be shown that the weights that minimize the reconstruction noise are given by,
\begin{equation}
\hat{w}^L_{l_1 l_2} = \frac{(H^L_{l_1 l_2})^2}{C_{l_1}C_{l_2}}\left[\sum_{l_1 l_2} \frac{(H^L_{l_1 l_2})^2}{C_{l_1}C_{l_2}} \right]^{-1} \,.
\label{eq:wght_fs_mve}
\end{equation}
Finally the reconstruction noise of the minimum variance estimator is given by,
\begin{equation}
N_L = \left[\sum_{l_1 l_2} \frac{(H^L_{l_1 l_2})^2}{2C_{l_1}C_{l_2}} \right]^{-1} \,.
\end{equation}
%


\section{Partial sky statistic}
\label{sec:psestmtr}

We follow a procedure similar to the one described in Appendix~\ref{sec:fsestmtr}, to arrive at a minimum variance estimator in the presence of a mask. Following Eq.~[\ref{eq:biposh:mask-unmask-aniso-approx}],
we can define a minimum variance estimator for any isotropy violating field $\beta(\hat{n})$ as,
\begin{equation}
\hat{\beta}_{LM} = \sum_{l_1 l_2} \hat{w}^{L}_{l_1 l_2} \frac{\hat{\mathcal{A}}^{LM}_{l_1 l_2}}{K^{LM}_{L\,M\,l_1\,l_2}} \,,
\label{eq:ps-estmr-lc}
\end{equation}
where $\hat{\mathcal{A}}^{LM}_{l_1l_2} = \hat{A}^{LM}_{l_1l_2} - \langle A^{LM}_{l_1 l_2} \rangle$, $\hat{A}^{LM}_{l_1l_2}$ are the BipoSH coefficients measured from the data, and $\langle A^{LM}_{l_1 l_2} \rangle$ denotes the bias due to all known systematic effects in the data like masking, anisotropic noise, non-circular beam, etc., estimated from simulations which incorporate all these systematic effects. It is non-trivial to evaluate the covariance of the BipoSH coefficients $\hat{\mathcal{A}}^{LM}_{l_1l_2}$ analytically, as can be done in the case of ideal full sky CMB. So we approximate the BipoSH covariance to be diagonal i.e.,
\begin{equation}
\langle \hat{\mathcal{A}}^{LM}_{l_1 l_2} \hat{\mathcal{A}}^{LM^*}_{l'_1 l'_2} \rangle_{Sim.} \approx \left(\hat{\sigma}^{LM}_{l_1 l_2}\right)^2  \delta_{l_1 l'_1} \delta_{l_2 l'_2} \delta_{LL'} \delta_{MM'} \,,
\end{equation}
where $\left(\hat{\sigma}^{LM}_{l_1 l_2}\right)^2$  is the variance of BipoSH coefficients which is estimated from simulations. Note that the covariance of BipoSH coefficients have an explicit $M$ dependence which was absent in the full sky BipoSH covariance given in Eq.~[\ref{biposh-cov}]. We also assume that the weights for the different BipoSH modes are independent of each other as in Eq.~[\ref{ind-w}]. Using these approximations and following the same procedure discussed in Appendix~\ref{sec:fsestmtr}, one finds that the weights that minimize the variance of the estimator are given by,
\begin{equation}
\hat{w}^{L}_{l_1 l_2} = \frac{1}{\sum_{M} \left(\frac{\hat{\sigma}^{LM}_{l_1 l_2}}{K^{LM}_{L M l_1 l_2}}\right)^2}
\left[
\sum_{l'_1 l'_2} \frac{1}{\sum_{M} \left(\frac{\hat{\sigma}^{LM}_{l'_1 l'_2}}{K^{LM}_{L M l'_1 l'_2}}\right)^2}
\right]^{-1} \,.
\end{equation}

Note that even though the weights themselves do not have an explicit $M$ dependence, the effective weighting of the BipoSH coefficients inferred from data is $M$ dependent owing to the presence of the modified shape function (see Eq. [\ref{eq:ps-estmr-lc}]).

\end{document}